\documentclass[article]{revtex4-1}
\usepackage{amsmath}
\usepackage{amssymb}
\usepackage{amsfonts}
\usepackage{amsthm}
\usepackage{amsbsy}
\usepackage[pdftex]{graphicx}										
\usepackage{color}
\usepackage{mathtools}
\usepackage{hyperref}
\usepackage{csquotes}
\usepackage{enumitem}
\usepackage{accents}

\theoremstyle{plain}
\newtheorem{thm}{Theorem}

\theoremstyle{remark}

\newcommand{\td}[2]{\frac{d{#1}}{d{#2}}}								
\renewcommand{\vec}[1]{\boldsymbol{\mathbf{#1}}}
\newcommand{\uvec}[1]{\vec{e}_{#1}}									
	
\DeclareMathOperator{\grad}{\nabla}

\newcommand\etal{\textit{et al}.\ }

\newcommand{\op}{\prime}
\newcommand{\tp}{{\prime\prime}}

\begin{document}
\title{Plane Poiseuille flow through a porous medium - an analytical solution}
\author{Amey S. Joshi}
\affiliation{Fidelity Investments, Embassy Golf Links Business Park, Bangalore, 560071}
\email{amey.joshi@outlook.com}
\date{\today}
\begin{abstract}
This paper develops a closed-form, analytical expression for the Darcy velocity of a Newtonian fluid flowing through a channel filled with a porous medium, bound by rigid walls and driven 
by a constant pressure gradient. We express the Darcy velocity as a Weierstrass elliptic function of the transverse coordinate. It allows us to get analytical expressions for volume flux 
and the rate of dissipation. The Weierstrass elliptic function is a doubly-periodic, complex-valued function defined over the complex plane. It takes real values only on certain segments 
in the complex plane. We show how to align the transverse coordinate axis along these segments to ensure real values of Darcy velocity. Lastly, we give an algorithm to generate velocity 
profile and compute volume flux given the flow parameters.
\end{abstract}

\maketitle

\section{Introduction}\label{s1}
Many flows occurring in nature are through channels filled with porous media. Some examples of such flows are ground water seeping through the earth, oil flowing through sand beds, blood 
flowing through vessels blocked by cholesterol and interstitial fluid flow in soft connective tissue. The need to study flow of fluids through porous media also arises in engineering
\cite{jambhekar2011forchheimer}. Catalytic converter for automobile exhaust system has the exhaust gases passing through a porous matrix of catalysts. Air cooled condensers pass hot air 
through porous sponges. Gas turbine blades have pores on them to allow cooling of the material of the blade. 

Quite like the rest of fluid mechanics, there are very few situations where an analytical solution can be found for given flow conditions. A plane Poiseuille flow with porous walls, across
which a fluid is injected or sucked with a constant, uniform velocity was first analyzed by Berman \cite{berman1953laminar} and later by Terill \cite{terrill1964laminar, 
terrill1965laminar}, Raithby \cite{raithby1971laminar}, Robinson \cite{robinson1976existence}, Skalak \etal\cite{skalak1978nonunique} and Shih\cite{shih1987existence}. We refer to Drazin 
and Riley's monograph\cite{drazin2006navier} for a few more examples of flow with porous walls. Vafai \etal\cite{vafai1990fluid} dealt with fluid mechanics at the interface between a fluid 
layer and a porous medium. They obtained an analytical solution for a flow over a flat, solid plate and a porous boundary above. Khan \etal\cite{khan2014exact} investigated analytical 
solutions in two dimensional flows involving porous boundaries. Analytical solutions are valuable, not for their rarity, but also, because they are reliable test cases for solutions 
obtained either numerically or through approximation techniques. An exact solution of plane Poiseuille flow, through a channel filled with a porous medium and driven by a constant pressure 
gradient, was first given by Nield \etal\cite{nield1996forced}. It is a formal solution, in the sense that it expresses the transverse coordinate $y$ as an elliptic integral in the Darcy 
velocity $\bar{u}$. They call it an exact solution because it can be evaluated numerically to any desired accuracy and with efforts far lesser than needed to solve the full partial 
differential equation. This paper develops an analytical solution in a conventional form, that is, expressing the Darcy velocity $\bar{u}$ as a function of the transverse coordinate $y$. 
This form allows us to get an expression for the flux and energy dissipation in the channel. Further, it is more suitable for a deeper analysis than the exact solution of Nield \etal.

In section \ref{s2}, we briefly introduce flows through porous media and set up the basic equations describing them. We define our problem in section \ref{s3} and obtain the 
equation of motion for it. We derive its solution in section \ref{s4}. It is an expression involving Weierstrass elliptic function and two constants of integration. In section \ref{s99}, 
we review the basic properties of Weierstrass elliptic function. These properties and the boundary conditions help us evaluate the constants of integration section in \ref{s5}. Section 
\ref{s6} has formulas for volume flux and rate of energy dissipation. In section \ref{s7}, we illustrate the theory developed so far, by plotting the velocity profile for a flow and 
calculating its volume flux. The algorithm to do so is described in appendix \ref{a2}. Weierstrass elliptic functions are no longer a part of an engineer's or a physicist's tool kit. 
Therefore, in appendix \ref{a1}, we briefly describe their properties and state a few useful theorems. We prove them in appendix \ref{a3}.

\section{Basic definitions and equations}\label{s2}
A porous medium is a material consisting of a fixed, solid matrix punctuated with interconnected channels. A fluid can flow through the channels but not through the solid matrix. The 
extent of availability of interconnected channels for a fluid to flow is characterized by the medium's porosity, $\varepsilon$. If $V_t$ is the total volume of a porous medium and $V_v$ is 
the volume of the interconnected channels in it then 
\begin{equation}\label{s2.e1}
\varepsilon = \frac{V_v}{V_t},
\end{equation}
Since the interconnected channels are a part of the medium, $V_v < V_t$, or equivalently, $0 < \varepsilon < 1$. Media with low values of $\varepsilon$ are called densely packed media, for 
example, $\epsilon \approx 0.02$ for coal. Media with values of $\varepsilon$ close to $1$ are called sparsely packed media, for example animal feathers, furs and high porosity metallic 
foams\cite{straughan2008stability}. 

Flow through a porous medium is a flow through the medium's interconnected channels. The channels have an irregular shape and orientation. Since the solid matrix is impervious to the 
fluid, the velocity field in a porous medium has non-zero values only in the interior of the channels. It is mathematically difficult to model a vector field taking non-zero values only in 
irregularly shaped portions of a channel. Therefore, we consider the average velocity of the fluid over a portion of the porous medium, sufficiently large to contain many channels and yet 
small comparable to the typical length scale of the flow. Such a portion is called a \enquote*{representative elementary volume} (REV). If the channels have spatial dimensions of the order 
of $d$ then we consider an REV in the form of a parallelepiped with dimensions large compared to $d$. The flux of fluid volume per unit area across the three orthogonal plane surfaces of 
the REV form the three components of a vector $\vec{\bar{u}}_{\star}$, called the Darcy velocity. The mechanical pressure field too, has complications like the velocity field. Therefore, 
we consider a similar averaging over an REV, and let $\bar{p}_{\star}$ describe the pressure field in the flow. We shall follow a convention of using a star subscript to denote variables 
with dimensions. Dimensionless variables, to be introduced in section \ref{s4}, will be denoted without a star subscript.

The first theory for a flow through a porous medium was proposed by Darcy\cite{darcy1856fontaines}. For an isotropic medium, it is described by the equation
\begin{equation}\label{s2.e2}
\grad\bar{p}_{\star} = -\frac{\mu}{K}\vec{\bar{u}}_{\star},
\end{equation}
now called Darcy's law. In this equation, $\mu$ is the fluid's viscosity and the parameter $K$ is the medium's permeability. $K$ is a function 
of the medium's porosity and the diameter, $d_p$, of solid particles and is usually defined as \cite{guo2002lattice},\cite{vafai1984convective}
\begin{equation}\label{s2.e3}
K = \frac{\varepsilon^3 d_p^2}{150(1 - \varepsilon)^2}
\end{equation}
It is a measure of ease of flow of a fluid through the medium. 

For a flow through sparsely packed media, where effects of the fluid boundary at solid particles are important, the Darcy law is extended by adding one more term, named after Brinkman, on 
the right hand side of \eqref{s2.e2} so that
\begin{equation}\label{s2.e4}
\grad\bar{p}_{\star} = -\frac{\mu}{K}\vec{\bar{u}}_{\star} + {\mu_e}\nabla^2\vec{\bar{u}}_{\star},
\end{equation}
where $\mu_e$ is the effective viscosity of the fluid. It is related to the fluid's dynamic viscosity $\mu$, in the case of an isotropic porous medium, by \cite{bear2012introduction}
\begin{equation}\label{s2.e5}
\frac{\mu_e}{\mu} = \frac{\lambda}{\epsilon},
\end{equation}
where $\lambda$, the tortuosity of the medium, is the ratio of the length of a stream line between two points and the straight line distance between the same points. 

When inertial effects are not negligible compared to the viscous effects, equation \eqref{s2.e4} has to be augmented with Forchheimer term \cite{khaled2003role}, so that
\begin{equation}\label{s2.e6}
\grad\bar{p}_{\star} = -\frac{\mu}{K}\vec{\bar{u}}_{\star} + {\mu_e}\nabla^2\vec{\bar{u}}_{\star} - 
\frac{\rho\varepsilon F_\varepsilon}{\sqrt{K}}|\vec{\bar{u}}_{\star}|\vec{\bar{u}}_{\star},
\end{equation}
where $F_\varepsilon$ is called the geometric function. Ergun's experiments \cite{ergun1952fluid} suggested that $F_\varepsilon$ depends only on the porosity of the medium as
\begin{equation}\label{s2.e7}
F_\varepsilon = \frac{1.75}{\sqrt{150\epsilon^3}}
\end{equation}
We refer to the paper by Khalid and Vafai\cite{khaled2003role} for a review of the three models, expressed as equations \eqref{s2.e2}, \eqref{s2.e4} and \eqref{s2.e6}, and a few examples of physical
situations in which they can be profitably used.

\section{The Problem Statement}\label{s3}
Consider a steady, fully-developed flow of a Newtonian fluid along the $x$ axis between two infinite parallel plates, a distance $H$ apart. Let the plates coincide with planes $y_{\star}
=0$ and $y_{\star}=H$. Let the flow be driven by a constant pressure gradient $\grad\bar p_{\star} = \rho\vec{G}_{\star}$. If the fluid is flowing parallel to the positive $x$-axis, let 
$\grad\bar{p}_{\star} = -\rho G_{\star}\uvec{x}$, where $\uvec{x}$ is the unit vector along the $x$-axis.  If the channel is long enough that the end effects are negligible along most of 
the channel's length then we can express the $x$ component of Darcy velocity as $\bar{u}_{\star}(y_{\star})$. Equation \eqref{s2.e6}, in this case, becomes 
\begin{equation}\label{s3.e1}
\frac{\nu_e}{\varepsilon}\bar{u}_{\star}^{\tp} - \frac{\nu}{\varepsilon K}\bar{u}_{\star} - \frac{F_\varepsilon}{\sqrt{K}}\bar{u}_{\star}^2 + \frac{G_{\star}}{\varepsilon} = 0,
\end{equation}
where $u^{\tp}_{\star}$ denotes the second derivative with respect to $y_\star$, $\nu_e = \mu_e/\rho$ is the effective kinematical viscosity and $\nu=\mu/\rho$ is the kinematic viscosity. 
Since the channel is bounded by solid walls, the usual no-slip boundary conditions apply (refer to equation (20) of Guo and Zhao's paper\cite{guo2002lattice}). We will solve \eqref{s3.e1} 
subject to the conditions $\bar{u}_{\star}(0) = 0$ and $\bar{u}_{\star}^\op(H/2) = 0$. The second of these conditions requires that the Darcy velocity has an extremum (actually, a maximum) 
at the center of the channel. 

\section{The solution}\label{s4}
We convert equation \eqref{s3.e1} in a non-dimensional form using the relations,
\begin{eqnarray}
\bar{u} &=& \frac{\bar{u}_{\star}}{U}		\label{s4.e1}	\\
y &=& \frac{y_{\star}}{H}		\label{s4.e2}	\\
G &=& \frac{HG_{\star}}{U^2},	\label{s4.e3}
\end{eqnarray}
where $U$ is the velocity at the center of the channel, and the dimensionless quantities
\begin{eqnarray}
Re &=& \frac{HU}{\nu}		\label{s4.e4}	\\
Da &=& \frac{K}{H^2}		\label{s4.e5}	\\
J  &=& \frac{\nu_e}{\nu}	\label{s4.e6}.
\end{eqnarray}
$Re$, $Da$ and $J$ are the Reynold number, the Darcy number and the viscosity ratio of the flow. Thus, equation \eqref{s3.e1} becomes,
\begin{equation}\label{s4.e7}
\bar{u}^{\tp} - A^2\bar{u}^2 - B^2\bar{u} + C^2 = 0,
\end{equation}
where
\begin{eqnarray}
A^2 &=& \frac{F_\varepsilon \varepsilon Re}{J\sqrt{Da}} \label{s4.e8} \\
B^2 &=& \frac{1}{J Da} \label{s4.e9} \\
C^2 &=& \frac{\varepsilon Re G}{J} \label{s4.e10}
\end{eqnarray}
are all positive numbers. Introduce a new variable $\bar{u}_1(y)$, 
\begin{equation}\label{s4.e11}
\bar{u}_1(y) = A\bar{u}(y) + \frac{1}{2}\frac{B^2}{A},
\end{equation}
so that equation \eqref{s4.e7} becomes,
\begin{equation}\label{s4.e12}
\bar{u}_1^\tp - A\bar{u}_1^2 + \left(AC^2 + \frac{B^4}{4A}\right) = 0,
\end{equation}
Multiplying equation \eqref{s4.e11} by $u_1^\op$, the first derivative of $\bar{u}_1(y)$ with respect to $y$, and integrating we get,
\begin{equation}\label{s4.e13}
\left({\bar{u}_1^\op}\right)^2 - \frac{2}{3}A\bar{u}_1^3 + 2\left(AC^2 + \frac{B^4}{4A}\right)\bar{u}_1 + \frac{36g_3}{A^2} = 0,
\end{equation}
where $36g_3/A^2$ is a constant of integration. 

Nield \etal rearranged \eqref{s4.e13} as
\begin{equation}\label{s4.e13b}
\left(\frac{2F}{3M}\right)^{1/2}\frac{dy}{d\bar{u}_1} = -\frac{1}{\{P(\bar{u}_1)\}^{1/2}},
\end{equation}
where $F$ and $M$ are constants and $P$ is a cubic polynomial in $\bar{u}_1$. They assumed that $P$ has real roots $b_1 < b_2 < b_3$ and that $b_2$ is the velocity at the center of the 
channel so that the first derivative vanishes there. Integrating equation \eqref{s4.e13b},
\[
\left(\frac{2F}{3M}\right)^{1/2}y = -\int\frac{d\bar{u}_1}{\{P(\bar{u}_1)\}^{1/2}} + c_0,
\]
where $c_0$ is a constant of integration. The first term on the right hand side is an elliptical integral in $\bar{u}_1$. The boundary conditions allow them to write
\begin{equation}\label{s4.e18}
y = \left(\frac{3M}{2F}\right)^{1/2}\frac{f(\phi \setminus \alpha)}{b_1 - b_3},
\end{equation}
where
\begin{eqnarray*}
\sin^2\alpha &=& \frac{b_2 - b_3}{b_1 - b_3} \\
\cos^2\alpha &=& \frac{(b_1 - b_2)(\bar{u}_1 - b_3)}{(b_2 - b_3)(b_1 - \bar{u}_1)}
\end{eqnarray*}
and $f(\phi \setminus \alpha)$ is an elliptic integral of first kind as defined by formula 17.4.63 in Abramowitz and Stegun\cite{abramowitz1964handbook}. Although a solution of the form 
$y$ equals an elliptic integral in $\bar{u}$ is exact in the sense meant by Nield \etal, it is not in a form convenient for a mathematical analysis. Therefore, we do not stop at 
equation \eqref{s4.e13}, but manipulate it further by introducing yet another variable
\begin{equation}\label{s4.e13a}
\bar{u}_2(y) = \frac{A}{6}\bar{u}_1(y),
\end{equation}
so that it becomes
\begin{equation}\label{s4.e14}
\left({\bar{u}_2^\op}\right)^2 = 4\bar{u}_2^3 - \frac{1}{3}\left(A^2C^2 + \frac{B^4}{4}\right)\bar{u}_2 - g_3.
\end{equation}
This equation is of the form,
\begin{equation}\label{s4.e15}
(\bar{u}^\op_2)^2 = 4\bar{u}_2^3 - g_2 \bar{u}_2 - g_3,
\end{equation}
where
\begin{equation}\label{s4.e16}
g_2 = \frac{1}{3}\left(A^2C^2 + \frac{B^4}{4}\right) 	
\end{equation}
Its solution is $\bar{u}_2(y) = \wp(y + k_1 ; g_2, g_3)$, where $k_1$ is another constant of integration. The function $\wp$ is the Weierstrass elliptic function with invariants 
$g_2$ and $g_3$ \cite{whittaker1996course}. Therefore, the general solution of \eqref{s4.e7} is
\begin{equation}\label{s4.e17}
\bar{u}(y) = \frac{6}{A^2}\left[\wp\left(y + k_1 \mathbin{;} \frac{1}{3}\left(A^2C^2 + \frac{B^4}{4}\right), g_3\right) - \frac{B^2}{12}\right]
\end{equation}
We will find the values of $g_3$ and $k_1$, the constants of integration, using the boundary conditions in section \ref{s5}. But before we do so, we will review the properties of 
Weierstrass elliptic function that will be of immediate use in the following sections.

\section{Weierstrass elliptic function}\label{s99}
Weierstrass elliptic function, $\wp$, is a doubly-periodic function defined over the complex plane. It is an even function\cite{lawden2013elliptic} of $z$. We follow the convention, 
preferred by Greenhill\cite{greenhill1892applications} and Lawden\cite{lawden2013elliptic}, to denote its two periods by $2\omega_1$ and $2\omega_3$. (An alternative convention, of 
denoting the periods by $2\omega_1$ and $2\omega_2$, is used by Whittaker and Watson\cite{whittaker1996course}.) Then for any integers $m$ and $n$,
\begin{equation}\label{s99.e1}
\wp(z + 2m\omega_1 + 2n\omega_3) = \wp(z),
\end{equation}
where $z$ is a complex number. The numbers $z$ and $z_{m,n} = z + 2m\omega_1 + 2n\omega_3$ are said to be congruent to each other. $\wp$ is finite throughout the complex plane except 
at its singularities. All singularities of $\wp$ are poles of order 2. They are at the origin and at all points congruent to it. The function $\wp$ also satisfies the differential 
equation
\begin{equation}\label{s99.e2}
\left[\wp^\op(z)\right]^2 = 4\wp^3(z) - g_2\wp(z) - g_3,
\end{equation}
where the constants $g_2$ and $g_3$ are called invariants of the Weierstrass elliptic function. In order to denote the dependence of value of $\wp$ on $z$ as well as the invariants, it is
sometimes written as $\wp(z \mathbin{;} g_2, g_3)$. If $2\omega_1$ and $2\omega_3$ are the two periods of $\wp$, then it is easy to see that $2\omega_2$ is also a period where
\begin{equation}\label{s99.e3}
\omega_2 = -\omega_1 - \omega_3
\end{equation}
If we define the constants 
\begin{equation}\label{s99.e4}
e_j = \wp(\omega_j),
\end{equation}
where $j = 1, 2, 3$, then it can be shown that\cite{lawden2013elliptic} $e_j$ are the three roots of the cubic equation
\begin{equation}\label{s99.e8}
4t^3 - g_2t - g_3 = 0
\end{equation}
whose discriminant is
\begin{equation}\label{s99.e9}
D = 16(g_2^3 - 27g_3^2)
\end{equation}
The nature of the roots, $e_1, e_2, e_3$, depends on the sign of $D$. If,
\begin{itemize}
\item $D > 0$, then all roots of \eqref{s99.e8} are real and distinct.
\item $D = 0$, then the roots of \eqref{s99.e8} are real and not distinct.
\item $D < 0$, then \eqref{s99.e8} has one real root and two complex roots. The complex roots are conjugates of each other.
\end{itemize}
Irrespective of the sign of the  discriminant, the constants $e_j$, are related to the invariants as
\begin{eqnarray}
g_2 &=& -4(e_1e_2 + e_2e_3 + e_3e_1) \label{s99.e5} \\
g_3 &=& 4e_1e_2e_3 \label{s99.e6}
\end{eqnarray}
and have he property
\begin{equation}\label{s99.e7}
e_1 + e_2 + e_3 = 0
\end{equation}
Weierstrass elliptic functions also appear in the solution of a two dimensional radial flow between two inclined walls. We refer the reader to Rosenhead's paper\cite{rosenhead1940steady} 
for a brief review of the properties of Weierstrass elliptic functions, useful in fluid dynamical problems.

\section{Finding \texorpdfstring{$g_3\text{ and }k_1$}{}}\label{s5}
We are now in a position to use the boundary conditions to find the constants of integration in the solution given by equation \eqref{s4.e17}. Since the Darcy velocity has an extremum at 
the center of the channel, its derivative with respect to $y$ vanishes there. Therefore, derivatives of $u_1$ and $u_2$ also vanish at the center of the channel. If the magnitude of the 
non-dimensional velocity $\bar{u}$ at the center is $u^c = 1$ then we immediately have $\bar{u}_1^c = A + B^2/(2A)$ and 
\begin{equation}\label{s5.e6}
\bar{u}_2^c = \frac{A^2}{6} + \frac{B^2}{12}
\end{equation}
Since at the center,
\[
4[\bar{u}_2^c]^3 - g_2 \bar{u}_2^c - g_3 = 0,
\]
$\bar{u}_2^c$ is one of the roots of the cubic equation $4s^3 - g_2s - g_3 = 0$. Since $e_1, e_2, e_3$ are its roots, without loss of generality, we let
\begin{equation}\label{s5.e7}
e_1 = \bar{u}_2^c
\end{equation}
Using equations \eqref{s99.e5} to \eqref{s99.e7} we readily get,
\begin{eqnarray}
e_2 &=& \frac{-e_1 + \sqrt{g_2 - 3e_1^2}}{2} \label{s5.e8} \\
e_3 &=& \frac{-e_1 - \sqrt{g_2 - 3e_1^2}}{2} \label{s5.e9},
\end{eqnarray}
From equations \eqref{s4.e16} and \eqref{s5.e6}, we observe that $g_2$ and $e_2$ depends only on the constants $A$ and $B$. Referring to equations \eqref{s4.e8} and \eqref{s4.e9}, we 
see that $A$ and $B$, depend only on the dimensionless numbers describing the flow. Thus, once $Re, Da$ and $J$ are specified, we get $g_2$ and $e_2$. From equations \eqref{s5.e8} and 
\eqref{s5.e9}, we also get $e_1$ and $e_3$. Knowing $e_1, e_2$ and $e_3$ equation \eqref{s99.e6} gives,
\begin{equation}\label{s5.e10}
g_3 = 4e_1e_2e_3
\end{equation}
We will now use the second boundary condition $\bar{u}(0) = 0$ to find $k_1$. The equation $\bar{u}(0) = 0$ is equivalent to, $\wp(0 + k_1 \mathbin{;} g_2, g_3) = B^2/12$ or
\begin{equation}\label{s5.e11}
k_1 = \wp^{-1}\left(\frac{B^2}{12} \mathbin{;} g_2, g_3\right)
\end{equation}
Thus, the particular solution of equation \eqref{s4.e7} is
\begin{equation}\label{s5.e12}
\bar{u}(y) = \frac{6}{A^2}\left[\wp\left(y + k_1 \mathbin{;} \frac{1}{3}\left(A^2C^2 + \frac{B^4}{4}\right), g_3\right) - \frac{B^2}{12}\right],
\end{equation}
where the constants of integration, $g_3$ and $k_1$, are found using equations \eqref{s5.e10} and \eqref{s5.e11}.

\section{Volume flux and rate of dissipation}\label{s6}
The flux of volume of fluid past a section of the channel is probably more important in practice than the velocity profile. If $y$ denotes the transverse dimension of the channel then the 
non-dimensional flux $Q$ is
\begin{equation}\label{s6.e1}
Q = \int_0^1 \bar{u}(y)dy,
\end{equation}
To get an analytical expression for flux in the case of a velocity field described by  \eqref{s5.e10}, \eqref{s5.e11} and \eqref{s5.e12}, we use the relationship between Weierstrass 
elliptic function and Weierstrass (\emph{not the same as Riemann}) zeta function \cite{lawden2013elliptic},
\begin{equation}\label{s6.e2}
\wp(y) = -\zeta^\op(y)
\end{equation}
From equations \eqref{s4.e17}, \eqref{s6.e1} and \eqref{s6.e2}, we get
\begin{equation}\label{s6.e3}
Q = \frac{6}{A^2}\left[\zeta(k_1 \mathbin{;} g_2, g_3) - \zeta(1 + k_1 \mathbin{;} g_2, g_3)\right] - \frac{B^2}{2A^2},
\end{equation}
the closed-form, analytical expression for volume flux. 

The non-dimensional rate of dissipation is
\[
\Phi = \left(\frac{\partial\bar{u}}{\partial y}\right)^2
\]
so that
\begin{equation}\label{s6.e4}
\Phi = \frac{36}{A^4}\left[\wp^\op\left(y + k_1 \mathbin{;} \frac{1}{3}\left(A^2C^2 + \frac{B^4}{4}\right), g_3\right)\right]^2
\end{equation}
The R programming environment provides a library \enquote*{elliptic}\cite{hankin2006introducing} to compute $\wp$, $\zeta$ and $\wp^\op$. In the next section, we report
results obtained by using this library. We used Mathematica\textsuperscript{\textregistered} to calculate inverse of Weierstrass elliptic function because the library 
\enquote*{elliptic} does not provide support for it. 

\section{Numerical results}\label{s7}
\begin{figure}[!ht]
\caption{Velocity profile of a flow in a porous medium}
\centering
\includegraphics[scale=1]{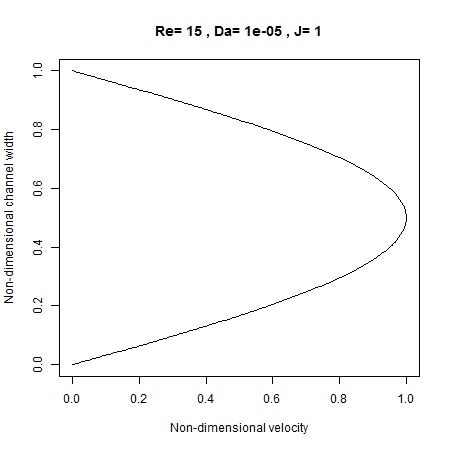}\label{f1}
\end{figure}

\begin{figure}[!ht]
\caption{Comparison of velocity profiles}
\centering
\includegraphics[scale=1]{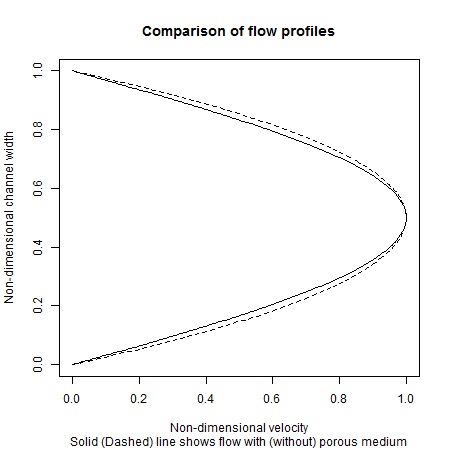}\label{f2}
\end{figure}

In this section we show how to use the analytical expressions obtained previously to plot the velocity profile and calculate the volume flux. We consider the flow of 
water through a channel of width $H = 0.1$ $m$ with a geometry as described in section \ref{s3}. The channel is filled with a porous medium of porosity 
$\varepsilon = 0.1$ and the flow is driven by a constant pressure gradient $G = 10^{-4}$ $Pa/m$. The dynamic viscosity of water is assumed to be $10^{-3}$ $Pa.s$ and the kinematic 
viscosity is assumed to be $10^{-6}$ $m^2/s$. We will consider a flow characterized by $Re = 15$, $Da = 10^{-5}$ and $J = 1$, where the Reynold number, Darcy number and viscosity ratio are 
defined in equations \eqref{s4.e4}, \eqref{s4.e5} and \eqref{s4.e6}. For this choice of flow parameters, the ratio of non-linear to linear drag\cite{guo2002lattice}, $(Re\sqrt{Da})^{-1}$, 
is approximately $21$. Therefore, the Forchheimer term in \eqref{s2.e6} cannot be neglected and we are justified in using the analytical results developed in this paper. The velocity 
profile of this flow is shown in figure \ref{f1}.
We now compare the shape of the velocity profile of a flow in a porous medium with a flow without the porous medium. If the pressure gradient for the two flows is identical, then the
velocity without porous medium is significantly larger than when porous medium is present. In order to compare the shapes of the two velocity profiles, we reduce the pressure gradient
of the flow without porous medium, keeping other parameters unchanged, so that the maximum velocity of the two is same. Figure \ref{f2} shows the two velocity profiles.
The non-dimensional volume flux of the flow through the channel filled with porous medium is $0.630$. It is slightly less than $0.667$, the flux for a velocity profile with same
peak velocity but without porous medium. Appendix \ref{a2} to this paper describes an algorithm to get the velocity profiles of figures \ref{f1} and \ref{f2}. In particular, we explain
how to align the transverse ($y$) axis in the complex plane to ensure real values for the Darcy velocity.

\section{Conclusion}\label{s8}
We extended the analysis of Nield \etal\cite{nield1996forced} to bring the equation of motion in a form that defines the Weierstrass elliptic function. We demonstrated how to align the
coordinate axis, transverse to the channel, in the complex plane so that the Weierstrass elliptic function takes real values. The analytical form of Darcy velocity in terms of Weierstrass 
elliptic function allowed us to get an expression for volume flux of the flow and the rate of dissipation. Since the Weierstrass elliptic function takes complex values in general, we 
need special care to ensure that it describes a physical quantity like the Darcy velocity. We stated a few properties of Weierstrass elliptic function to help us achieve our goal. Finally,
in appendix \ref{a2}, we describe an algorithm to get velocity profile and calculate the volume flux.

\appendix

\section{Some more properties of Weierstrass elliptic function}\label{a1}
Weierstrass elliptic function, $\wp$, takes complex values over the complex plane. However, we want the Darcy velocity, $\bar{u}$, to be real. Therefore, we find out conditions under 
which $\wp$ takes real values. To that end, we state the following theorem and prove it in appendix \ref{a3}.
\begin{thm}\label{t1}
If the invariants $g_2$ and $g_3$ of Weierstrass elliptic function $\wp$ are real and if the roots of the cubic $4s^3 - g_2s - g_3 = 0$ are real then $\omega_1$ is real and $\omega_3$ 
is imaginary. 
\end{thm}
The converse of theorem \ref{t1} is also true [refer to p. 163 of Lawden\cite{lawden2013elliptic}]. That is, if $\omega_1$ is real and $\omega_3$ is imaginary then the roots, 
$e_1, e_2, e_3$, of $4s^3 - g_2s - g_3 = 0$ are all real. Therefore, by \eqref{s99.e5} and \eqref{s99.e6}, the invariants $g_2$ and $g_3$ are real.

\begin{figure}[!ht]
\caption{Positive discriminant}
\centering
\includegraphics[scale=1]{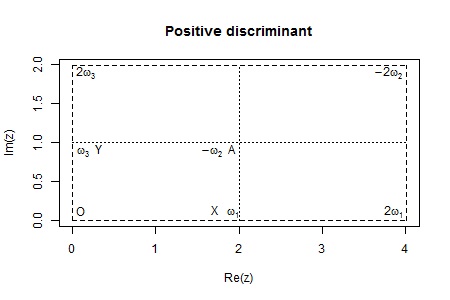}\label{f3}
\end{figure}

The double periodicity of $\wp$ allows us to choose $\omega_1$ and $\omega_3$ to be in the first quadrant of the complex plane. If $\omega_1$ is real and $\omega_3$ is imaginary then 
Lawden\cite{lawden2013elliptic} proved
\begin{thm}\label{t2}
$\wp$ takes real values on the the rectangle $OXAY$ in figure \ref{f3}, where $O \equiv (0, 0), X \equiv (\omega_1, 0), A \equiv (\omega_1, \omega_3)$ and $Y \equiv (0, \omega_3)$. 
Further, if $z$ is taken round $OXAY$, $\wp(u)$ decreases from $+\infty$ at $O$ to $e_1$ at $X$, further decreases to $e_2$ along $XA$, to $e_3$ along $AY$ and finally to $-\infty$ as it 
reaches the origin along $YA$. 
\end{thm}
Recall that, in equation \eqref{s5.e7}, we chose $e_1$ to be $\bar{u}_2^c$, the value of $\bar{u}_2$ at the center of the channel. On the other hand, $k_1$, was the value of $\bar{u}_2$
at the walls. We thus have two possibilities (i) $e_1 > k_1 > e_2 > e_3$, so that $k_1$ lies on segment $XA$ in figure \ref{f3} or (ii) $e_1 > e_2 > k_1 > e_3$, so that $k_1$ lies on 
segment $AY$ of figure \ref{f3}. If we denote the point representing $k_1$ by $B$ then segment $XB$ in the first case and $AB$ in the second case represents the half width of the channel.  
In non-dimensional quantities, we expect the half-width to be exactly $1/2$. However, in general, the length of the segment $XB$ or $AB$ is not half, requiring us to rescale the transverse 
coordinate once more. It is easy to prove
\begin{thm}\label{t3}
If we introduce a new non-dimensional displacement variable $\tilde{y}$ defined as 
\begin{equation}\label{a1.e1}
\tilde{y} = \frac{y}{\alpha},
\end{equation}
where $\alpha$ is a real number then this transformation keeps the form of equation \eqref{s4.e7} unchanged. Further, the new constants $\tilde{A}, \tilde{B}$ and $\tilde{C}$ are related
to the older constants $A$, $B$ and $C$ by
\begin{eqnarray}
\tilde{A} &=& \alpha A \label{a1.e2} \\
\tilde{B} &=& \alpha B \label{a1.e3} \\
\tilde{C} &=& \alpha C \label{a1.e4}
\end{eqnarray}
\end{thm}

\begin{figure}[!ht]
\caption{Negative discriminant}
\centering
\includegraphics[scale=1]{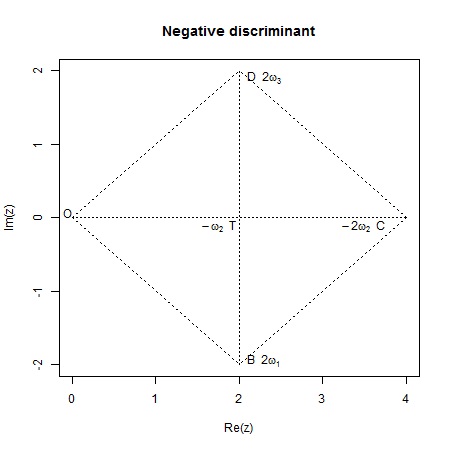}\label{f4}
\end{figure}

If the invariants $g_2$ and $g_3$ of the Weierstrass elliptic function are real but the discriminant of the cubic equation $4t^3 - g_2t - g_3 = 0$ is negative, then referring to figure 
\ref{f4} [or p. 630 of Abramowitz and Stegun\cite{abramowitz1964handbook}], the function $\wp$ takes value $+\infty$ at $O$. Its value decreases monotonically up to $-\omega_2$, where it 
is $e_2$ and then rises once again to $+\infty$ at $-2\omega_2$. If the function has a singularity at the origin, it will have it at points $-2\omega_2$, $2\omega_1$ and $2\omega_3$ 
because of its periodicity. The function also takes real values along the segment $BD$ where it decreases from $e_2$ to $-\infty$, both above and below. Since $e_2$ is the only real root
of $4t^3 - g_2t - g_3 = 0$, we choose $\bar{u}_2^c$ to be $e_2$ instead of $e_1$. Therefore, the transverse axis has to be aligned along $BD$. As in the case of positive discriminant, 
here too we have to rescale the variable $y$ and theorem \ref{t3} guides us in that procedure.

If the invariants $g_2$ and $g_3$ of the Weierstrass elliptic function are real but the discriminant of the cubic equation $4t^3 - g_2t - g_3 = 0$ is zero, then we have, what is usually
called\cite{rosenhead1940steady} the degenerate case. If $e_1, e_2, e_3$ are the roots of $4t^3 - g_2t - g_3 = 0$ then it is always true that $e_1 + e_2 + e_3 = 0$. Further, at least
two of the roots are identical. We therefore have three sub-cases
\begin{enumerate}
\item[case (i)] All roots are equal. Therefore, each one of zero. Since we chose $\bar{u}_2^c = e_1$, we have $\bar{u}_2^c = 0$ and hence $\bar{u}^c < 0$. Thus, if all roots are equal 
then the fluid velocity will be negative at the center of the channel. Therefore, we exclude this possibility on physical grounds.

\item[case (ii)] $e_1 > 0$ and $e_2 = e_3 = -e_1/2$. Since we chose the $\bar{u}_2^c = e_1$, the value of $\bar{u}_2$ at the walls has to be less then $e_1$. In particular, 
$\bar{u}_2(0) < e_1$. In this case, equation \eqref{s99.e2} can be written as $\left[\bar{u}_2^\op(y)\right]^2 = 
(\bar{u}_2(y) - e_1)(\bar{u}_2(y) + 2e_1)^2$. Therefore, $\left[\bar{u}_2^\op(0)\right]^2 < 0$, and hence $\left[\bar{u}^\op\right]^2 < 0$. Since Darcy velocity is a real quantity, it 
cannot have an imaginary derivative. Therefore, we exclude this possibility as well on physical grounds.

\item[case (iii)] $e_1 > 0$ and $e_1 = e_2 = -e_3/2$. Since the sum of the roots is always zero, we have $e_3 < 0$. In this case, equation \eqref{s99.e2} can be written as 
$\left[\bar{u}_2^\op(y)\right]^2 = (\bar{u}_2(y) - e_1)^2(\bar{u}_2(y) + 2e_1)$. The right hand side of this equation is always positive. Therefore, if the discriminant of the cubic 
equation $4t^3 - g_2t - g_3 = 0$ is zero then $e_1 > 0$ and $e_1 = e_2 = -e_3/2$ is the only possibility. The equation of $\bar{u}_2$ is thus 
$\left[\bar{u}_2^\op\right]^2 = (\bar{u}_2(y) - e_1)^2(\bar{u}_2(y) + 2e_1)$ or $\bar{u}_2^\op = -(\bar{u}_2(y) - e_1)\{\bar{u}_2(y) + 2e_1\}^{1/2}$, where we emphasize that the negative
sign of the square root is appropriate (refer to Lawden's\cite{lawden2013elliptic} equation (6.7.16)). Therefore,
\[
\int\frac{d\bar{u}_2}{(\bar{u}_2(y) - e_1)\{\bar{u}_2(y) + 2e_1\}^{1/2}} = - y - \frac{2\beta}{\sqrt{3e_1}},
\]
where $\beta$ is a constant of integration. The integral on the left hand side can be readily evaluated (by substituting for $\{\bar{u}_2(y) + 2e_1\}^{1/2}$) so that
\[
\tanh^{-1}\left(\frac{2}{3} + \frac{\bar{u}_2}{3e_1}\right) = \frac{\sqrt{3e_1}}{2}y + \beta
\]
or,
\begin{equation}\label{a1.e5}
\bar{u}_2 = 3e_1\tanh\left(\frac{\sqrt{3e_1}}{2}y + \beta\right) - 2e_1
\end{equation}
Using the boundary condition $\bar{u}_2(0) = B^2/12$, we get
\begin{equation}\label{a1.e6}
\beta = \tanh^{-1}\left(\frac{2}{3} + \frac{B^2}{36e_1}\right)
\end{equation}
Using the relations between $\bar{u}$, $\bar{u}_1$ and $\bar{u}_2$ given in equations \eqref{s4.e11} and \eqref{s4.e13a} we get the Darcy velocity, 
\begin{equation}\label{a1.e7}
\bar{u}(y) = \frac{18e_1}{A^2}\tanh\left(\frac{\sqrt{3e_1}}{2}y + \beta\right) - \frac{24e_1 + B^2}{2A^2}
\end{equation}
This form of the solution is usually reported as an analytical solution for a plane Poiseuille form in porous medium\cite{vafai1989forced,nield1996forced} and the related problem of a 
radial flow in between two inclined plane walls\cite{rosenhead1940steady}.
channel
\end{enumerate}

\section{An algorithm to compute velocity profile and flux}\label{a2}
In this section, we describe an algorithm to generate the figures \ref{f1} and \ref{f2} in section \ref{s7}.
\begin{enumerate}
\item\label{step1} Set the physical properties of the fluid ($\mu$, the dynamic viscosity and $\nu$, the kinematic viscosity) and the channel (width $H$, porosity $\varepsilon$ and 
pressure gradient $G$).

\item\label{step2} Choose the flow characteristics, that is Reynold number $Re$, Darcy number $Da$ and viscosity ratio $J$.

\item\label{step3} Calculate the geometric function using \eqref{s2.e7} and the non-dimensional pressure gradient using \eqref{s4.e3}.

\item\label{step4} Let scale $s = 1$ and calculate constants $A$, $B$ and $C$ using a slight modification of \eqref{s4.e8}, \eqref{s4.e9} and \eqref{s4.e10}
\begin{eqnarray}
A^2 &=& \frac{F_\varepsilon \varepsilon Re}{J\sqrt{Da}} \times s \label{a2.e1} \\
B^2 &=& \frac{1}{J Da} \times s \label{a2.e2} \\
C^2 &=& \frac{\varepsilon Re G}{J} \times s \label{a2.e3}
\end{eqnarray}
With $s = 1$, they are identical to their previous definitions. We will explain the reason for this additional variable in step \ref{step6}. Calculate the invariants $g_2$ using 
\eqref{s4.e16} and $g_3$ using \eqref{s5.e10}. Calculate the discriminant $D$ of the cubic $4t^3 - g_2t - g_3 = 0$ using \eqref{s99.e9}

\item\label{step5} Compute $k_1$ using \eqref{s5.e11}. The library \enquote*{elliptic} does not have facility to compute $\wp^{-1}$. We, therefore, used the 
Mathematica\textsuperscript{\textregistered} function \texttt{InverseWeierstrassP}.

\item\label{step6} In this step, we calculate the scale appropriate for non-dimensional transverse coordinate. If $D$ is positive, choose $\omega_1$, $\omega_3$ and $-\omega_2$ to be all 
in the first quadrant, as shown in figure \ref{f3}. If $\Im(z)$ is the imaginary part of the complex number $z$, then set scale $s = 2\Im(k_1)$ if $k_1$ lies on segment $XA$. If $k_1$ 
lies on $AY$, set $s = (\Im(-\omega_2) - \Im(k_1))$. If $D$ is negative, choose $\omega_1$, $\omega_2$ and $\omega_3$ as in figure \ref{f4}. Set scale $s = 2\Im(k1)$. If $s \ne 1$, go to 
step \ref{step4}, else proceed to step \ref{step7}.

\item\label{step7} Calculate the Darcy velocity over the half-width of the channel. The half-width was identified in step \ref{step6}.

\item\label{step8} The Darcy velocity of the other half of the channel is found by symmetry. Once we get the velocity over the entire channel, we can plot it against $[0, 1]$ to get 
figure \ref{f1}.

\item\label{step9} Find the maximum Darcy velocity in the channel. Our choice of non-dimensional variables gives the maximum Darcy velocity $1$. The pressure gradient giving this velocity
in a channel without porous medium, is $G_{eq} = -8/Re$. Plotting velocity profile for plane Poiseuille flow with pressure gradient $G_{eq}$ over that of the Darcy velocity gives figure 
\ref{f2}.

\item\label{step10} Find the non-dimensional flux in the case of channel filled with porous medium using \eqref{s6.e3}. The corresponding value for plane Poiseuille flow without porous
medium is $2/3$.
\end{enumerate}

\section{Proof of theorems \ref{t1} and \ref{t3}}\label{a3}
\begin{proof}[Proof of theorem \ref{t1}]
The constants $e_1, e_2, e_3$ are solutions of the cubic $4s^3 - g_2s - g_3 = 0$. If they are all real, we can choose them to be such that $e_1 > e_2 > e_3$. Then \eqref{s99.e2} can be
written as
\[
\left\{(\wp(z) - e_1)(\wp(z) - e_2)(\wp(z) - e_3)\right\}^{1/2} = -\wp^\op(z)
\]
that is,
\[
\frac{d\wp(z)}{\{(\wp(z) - e_1)(\wp(z) - e_2)(\wp(z) - e_3)\}^{1/2}} = -dz
\]
equivalently,
\[
\frac{dt}{\{4(t - e_1)(t - e_2)(t - e_3)\}^{1/2}} = -dz,
\]
where $t = \wp(u)$. Integrating this equation so that $z$ goes from $0$ to $U$, and hence $t$ goes from $\wp(0)$ to $\wp(U)$, we have two possibilities because of the double pole of
$\wp$ at the origin,
\begin{equation}\label{a3.e1}
\int_{\infty}^{\wp(U)}\frac{dt}{\{4(t - e_1)(t - e_2)(t - e_3)\}^{1/2}} = -U
\end{equation}
or
\begin{equation}\label{a3.e2}
\int_{-\infty}^{\wp(U)}\frac{dt}{\{4(t - e_1)(t - e_2)(t - e_3)\}^{1/2}} = -U
\end{equation}
Choose $U = \omega_1$ in equation \eqref{a3.e1}, so that
\[
\omega_1 = -\int_{\infty}^{\wp(\omega_1)}\frac{dt}{\{4(t - e_1)(t - e_2)(t - e_3)\}^{1/2}} = \int_{e_1}^{\infty}\frac{dt}{\{4(t - e_1)(t - e_2)(t - e_3)\}^{1/2}}
\]
Since $e_1$ is the largest root of $4(t - e_1)(t - e_2)(t - e_3) = 4t^3 - g_2t - g_3 = 0$, throughout the path of integration, the integrand stays finite. Further, the integrand is always
real and hence $\omega_1$ is real. 

Choose $U = -\omega_3$ in equation \eqref{a3.e2}, so that
\[
\omega_3 = -\int_{-\infty}^{\wp(\omega_3)}\frac{dt}{\{4(t - e_1)(t - e_2)(t - e_3)\}^{1/2}} = -\int_{-\infty}^{e_3}\frac{dt}{\{4(t - e_1)(t - e_2)(t - e_3)\}^{1/2}}
\]
or,
\[
\omega_3 = i\int_{-\infty}^{e_3}\frac{dt}{\{4(e_1 - t)(e_2 - t)(e_3 - t)\}^{1/2}}
\]
Since $e_3$ is the smallest root of $4(t - e_1)(t - e_2)(t - e_3) = 4t^3 - g_2t - g_3 = 0$, throughout the path of integration, the integrand stays finite. Further, the integrand is 
always real and hence $\omega_3$ is imaginary. 
\end{proof}

\begin{proof}[Proof of theorem \ref{t3}]
For sake of clarity, let us write the derivatives in full. Thus, equation \eqref{s4.e7} is written as
\begin{equation}\label{a3.e3}
\frac{d^2\bar{u}}{dy^2} - A^2\bar{u}^2 - B^2\bar{u} + C^2 = 0
\end{equation}
If the transverse dimension $y_\star$ is converted to a non-dimensional form $\tilde{y} = y_\star/(\alpha H)$, where $\alpha$ is a positive real number then
\[
\tilde{y} = \frac{y_\star}{\alpha H} = \frac{y}{\alpha}
\]
If $f$ is a function of $y$ then
\[
\td{f}{y} = \td{f}{\tilde{y}}\td{\tilde{y}}{y} = \frac{1}{\alpha}\td{f}{\tilde{y}}
\]
and hence
\[
\frac{d^2\bar{u}}{dy^2} = \frac{1}{\alpha^2}\frac{d^2 \bar{u}}{d\tilde{y}^2}
\]
If we denote the new dimensionless numbers with a tilde on top then from equations \eqref{s4.e4} to \eqref{s4.e6},
\begin{eqnarray}
\tilde{Re} &=& \alpha Re	\label{a3.e4}	\\
\tilde{Da} &=& \frac{Da}{\alpha^2}	\label{a3.e5}	\\
\tilde{J}  &=& J\label{a3.e6} 
\end{eqnarray}
The new pressure gradient is $\tilde{G} = \alpha G$ while the new constants $\tilde{A}$, $\tilde{B}$ and $\tilde{C}$ become
\begin{eqnarray}
\tilde{A}^2 &=& \frac{F_0\epsilon \tilde{Re}}{\tilde{J}\sqrt{\tilde{Da}}}  = \alpha^2 A^2 \label{a3.e7} \\
\tilde{B}^2 &=& \frac{\epsilon}{\tilde{J} \tilde{Da}}  = \alpha^2 B^2 \label{a3.e8} \\
\tilde{C}^2 &=& \frac{\epsilon^2 \tilde{Re}\tilde{G}}{\tilde{J}} = \alpha^2 C^2 \label{a3.e9}
\end{eqnarray}
Therefore, equation \eqref{a3.e3} becomes,
\[
\frac{1}{\alpha^2}\frac{d^2 \bar{u}}{d\tilde{y}^2} - \frac{\tilde{A}^2}{\alpha^2}\bar{u}^2 - \frac{\tilde{B}^2}{\alpha^2}\bar{u} + \frac{\tilde{C}^2}{\alpha^2} = 0
\]
or
\[
\frac{d^2 \bar{u}}{d\tilde{y}^2} - \tilde{A}^2\bar{u}^2 - \tilde{B}^2\bar{u} + \tilde{C}^2 = 0
\]
\end{proof}
\bibliography{ppfpm}
\end{document}